\documentclass[preprint,showpacs,preprintnumbers,amsmath,amssymb]{revtex4}

\usepackage{graphicx}
\usepackage{dcolumn}
\usepackage{bm}

\begin{document}

\noindent

\preprint{}

\title{Possible corrections to quantum mechanical predictions\\ in hidden variable model}

\author{Agung Budiyono}

\affiliation{Institute for the Physical and Chemical Research, RIKEN, 2-1 Hirosawa, Wako-shi, Saitama 351-0198, Japan}

\date{\today}

\begin{abstract} 

We derive possible corrections to the statistical predictions of quantum mechanics in measurement over ensemble of identically prepared system based on a hidden variable model of quantization developed in the previous work. The corrections are characterized by a dimensionless parameter $\sigma$ and the prediction of quantum mechanics is reproduced in the formal limit $\sigma\rightarrow 0$. Quantum mechanics is argued to be reliable for sufficiently low quantum number.  

\end{abstract}

\pacs{03.65.Ta, 03.65.Ca}
\keywords{hidden variable model; correction to quantum mechanical predictions}
\maketitle  

\section{Introduction}

For almost nine decades since its completion, quantum mechanics has been claimed to be confirmed by a wealth of experimental tests with unparalleled accuracy. Nevertheless, in view of its operational character \cite{Peres book}, the yet unsettled foundational problems \cite{Isham book}, and the difficulties in unifying quantum mechanics with general relativity \cite{Kiefer quantum gravity}, it is imperative to ask whether quantum mechanics is an accurate approximation of a deeper theory \cite{Bialynicki,Shimony,Ellis,Weinberg,Bertolami,Gamboa,tHooft,Markopoulou,Smolin,Adler,Perez-Sahlamnn-Sudarsky,Valentini,Groessing}. There are at least two possible directions to exercise this question. First, one keeps the formalism of quantum mechanics unchanged as far as possible and speculates a small correction to the fundamental equation of the theory. The other direction is to construct a conceptually new theory which reproduces the empirical statistical prediction of quantum mechanics as certain limiting case. In this latter approach, the operational formalism of quantum mechanics should be shown as emergent. Obviously, this approach is preferable to attack the foundational problems of quantum mechanics and the problems of quantum gravity. 

On the other hand, in our previous work \cite{AgungDQM1}, we have proposed a hidden variable model for quantization by modifying the classical dynamics of ensemble of trajectories parameterized by a hidden random variable. We showed for a wide class of important dynamical systems that, given the classical Hamiltonian, the modified equations can be put into the Schr\"odinger equation with a unique Hermitian quantum Hamiltonian and Born's statistical interpretation of wave function, if the probability density (mass) function of the hidden variable $\lambda$ takes a specific form as 
\begin{equation}
P_Q(\lambda;\hbar)=\frac{1}{2}\delta(\lambda-\hbar)+\frac{1}{2}\delta(\lambda+\hbar). 
\label{quantum condition}
\end{equation}  
Namely, $\lambda$ is an unbiased binary random variable which can only take values $\pm\hbar$. We have also shown that in this case one can always identify an ``effective'' velocity field which numerically is equal to the ``actual'' velocity field of particle in pilot-wave theory \cite{Bohm paper}. This then allows us to follow the description of pilot-wave theory on measurement without wave function collapse. Yet unlike pilot-wave theory, our model is inherently stochastic and the wave function is not physically real field. 
 
In the present paper, we shall further elaborate the hidden variable model of Ref. \cite{AgungDQM1} in the case when the distribution of $\lambda$ is allowed to deviate from Eq. (\ref{quantum condition}) while satisfies the following general condition:
\begin{eqnarray}
P(\lambda;\hbar,\sigma)=P(-\lambda;\hbar,\sigma),\hspace{10mm}\nonumber\\
\mbox{with}\hspace{2mm}\lambda\neq 0,\hspace{2mm}\mbox{and}\hspace{2mm}\lim_{\sigma\rightarrow 0}P(\lambda;\hbar,\sigma)=P_Q(\lambda;\hbar). 
\label{sub-quantum condition}
\end{eqnarray}
Hence, $\lambda$ is non-vanishing ($P(0;\hbar,\sigma)=0$), its distribution function is even so that it is unbiased, and we introduce a new parameter $\sigma$ so that Eq. (\ref{quantum condition}) is recovered as a formal limit $\sigma\rightarrow 0$. This will then be shown to offer possible corrections to the prediction of quantum mechanics in measurement over ensemble of identically prepared system. The discussion will be facilitated by considering concrete models of  measurement of angular momentum. First, we shall show that there is an inherent broadening of spectral line which is purely induced by the distribution of the hidden variable. Accordingly, the Born's statistical rule will also be shown to have small correction characterized by $\sigma$. 

\section{Hidden variable model for quantization: Quantum Hamiltonian for angular momentum measurement}    

Let us consider the dynamics of $N$ particles with configuration coordinate $q=(q_1,q_2,\dots,q_N)$. The classical Hamiltonian is denoted by $\underline{H}(q,\underline{p};t)$, where $\{{\underline{p}}_i\}$ is momentum canonically conjugate to $\{q_i\}$ and $t$ is time. All mathematical symbols with ``underline'' will be used to denote physical quantities satisfying the law of classical mechanics. The classical dynamics of the particles then follows the Hamilton-Jacobi equation
\begin{equation}
\partial_t\underline{S}(q;t)+\underline{H}(q,\partial_q\underline{S}(q;t);t)=0,
\label{H-J equation}
\end{equation}
where $\underline{S}(q;t)$ is the Hamilton principle function (HPF) so that $\underline{p}=\partial_q\underline{S}$ \cite{Rund book}. To solve the above equation, one needs to set up an initial HPF $\underline{S}(q;0)$ which implies an initial classical momentum field $\underline{p}(q;0)=\partial_q\underline{S}(q;0)$. A single trajectory in configuration space is picked up if one also fixes the initial position of the particles.

Let us then consider an ensemble of classical system so that the probability density function of the position of the particles  is denoted by $\underline{\rho}(q;t)$. It must satisfy the following continuity equation: 
\begin{equation}
\partial_t\underline{\rho}+\partial_{q}\cdot(\underline{v}(\underline{S})\underline{\rho})=0,
\label{continuity equation}
\end{equation}   
where $\underline{v}=(\underline{v}_1,\dots,\underline{v}_N)$ is the classical velocity field. In the above equation we have made explicit the possible dependence of the classical velocity field $\underline{v}$ on the HPF $\underline{S}$, which can be obtained from the classical Hamiltonian through the Hamilton equation: 
\begin{equation}
{\underline{v}}_i=\frac{\partial \underline{H}}{\partial {\underline{p}}_i}\Big|_{\{{\underline{p}}_i=\partial_{q_i}\underline{S}\}}=f_i(\underline{S}),
\label{classical velocity field}
\end{equation}
where $f_i$, $i=1,\dots,N$ are some functions determined by the choice of $\underline{H}$ as above \cite{note on constrained dynamics}. The dynamics and statistics of the ensemble of classical trajectories are then obtained by solving Eqs. (\ref{H-J equation}), (\ref{continuity equation}) and (\ref{classical velocity field}) in term of  $\underline{S}(q;t)$, $\underline{\rho}(q;t)$ and $\underline{v}(q;t)$.  

Now let us develop a general scheme to modify the above classical dynamics of ensemble of trajectories \cite{AgungDQM1}. To do this let us introduce two real-valued functions $S(q,\lambda;t,\hbar,\sigma)$ and $\Omega(q,\lambda;t,\hbar,\sigma)$, where $\lambda$ is a hidden random variable whose probability density function is assumed to satisfy Eq. (\ref{sub-quantum condition}). They are supposed to take over the role of $\underline{S}$ and $\underline{\rho}$ in the modified dynamics. Hence, $\Omega(q,\lambda;t,\hbar,\sigma)$ is the joint-probability density that the particles are at configuration space $q$ and the value of hidden random  variable is $\lambda$. The marginal probability densities of the fluctuations of $q$ and $\lambda$ are thus given, respectively, by 
\begin{eqnarray}
\rho(q;t,\hbar,\sigma)=\int d\lambda\Omega(q,\lambda;t,\hbar,\sigma),\nonumber\\
P(\lambda;\hbar,\sigma)=\int dq\Omega(q,\lambda;t,\hbar,\sigma).  
\label{marginal probabilities}
\end{eqnarray}
In the following, for notational simplicity, we shall sometime not make explicit notationally the dependence of any functions on $\hbar$ and $\sigma$.  

Next, let us proceed to assume the following general rule of replacement to modify Eqs. (\ref{H-J equation}) and (\ref{continuity equation}) \cite{AgungDQM1}: 
\begin{eqnarray}
\underline{\rho}\mapsto\Omega,\hspace{35mm}\nonumber\\
\partial_{q_i}\underline{S}\mapsto\partial_{q_i}S+\frac{\lambda}{2}\frac{\partial_{q_i}\Omega}{\Omega},\hspace{2mm}i=1,\dots,N,\nonumber\\
\partial_{t}\underline{S}\mapsto\partial_tS+\frac{\lambda}{2}\frac{\partial_t\Omega}{\Omega}+\frac{\lambda}{2}\partial_q\cdot f(S),\hspace{15mm}
\label{fundamental equation}
\end{eqnarray} 
where the vector-valued function $f=(f_1,\dots,f_N)$ is defined by Eq. (\ref{classical velocity field}). Let us first show that the replacement of Eq. (\ref{fundamental equation}) possesses a consistent classical correspondence if $S\rightarrow\underline{S}$ so that the Hamilton-Jacobi equation of (\ref{H-J equation}) is restored (notice that we have used the symbol ``$\mapsto$'' to denote replacement and ``$\rightarrow$'' to denote a limit). First, using the last two equations of (\ref{fundamental equation}), for sufficiently small $\Delta t$ and $\Delta q=\{\Delta q_i\}$, then expanding $\Delta F\doteq F(q+\Delta q;t+\Delta t)-F(q;t)\approx\partial_tF\Delta t+\partial_qF\cdot\Delta q$, for any function $F$, one has
\begin{equation}
\Delta\underline{S}\mapsto\Delta S+\frac{\lambda}{2}\Big(\frac{\Delta\Omega}{\Omega}+\partial_q\cdot f(S)\Delta t\Big). 
\label{statistical violation of HPSA}
\end{equation}  
One can see that in the limit $S\rightarrow\underline{S}$, in order to be consistent then the second term on the right hand side has to be vanishing. Namely one has $d\Omega/dt=-\Omega\partial_q\cdot\underline{v}$, by Eq. (\ref{classical velocity field}). This is just the continuity equation of (\ref{continuity equation}). In other words, in the limit $S\rightarrow\underline{S}$, $\rho(q;t)=\int d\lambda\Omega$ has to approach $\underline{\rho}$, $\rho\rightarrow\underline{\rho}$. 

Let us apply the above modification of classical mechanics to a class of von Neumann model of measurement of angular momentum. A different model of measurement will be given in Section \ref{Stern-Gerlach experiment}. To do this, let us consider the dynamics of two interacting particles with coordinate $q=(q_1,q_2)$. The first particle represents the system whose angular momentum is being measured and the second particle represents the measuring apparatus. To make explicit the three dimensional nature of the problem, let us put  $q_1=(x_1,y_1,z_1)$. For simplicity let us consider the measurement of $z-$part angular momentum of the first particle 
\begin{equation}
\underline{L}_{z_1}=x_1{\underline{p}}_{y_1}-y_1{\underline{p}}_{x_1}, 
\label{classical angular momentum}
\end{equation}
where ${\underline{p}}_{x_1}$ is the conjugate momentum of $x_1$ and so on. 

Next, let us choose the following measurement-interaction classical Hamiltonian:
\begin{equation}
\underline{H}_l=g\underline{L}_{z_1}\underline{p}_2=g(x_1{\underline{p}}_{y_1}-y_1{\underline{p}}_{x_1}){\underline{p}}_2,
\label{classical Hamiltonian angular momentum}
\end{equation}
where $g$ is the coupling parameter. Let us assume that the interaction is impulsive so that one can ignore the free Hamiltonian of each particles. First, $\underline{L}_{z_1}$ is time-invariant: $d\underline{L}_{z_1}/dt=\{\underline{L}_{z_1},\underline{H}_l\}=0$ where $\{\cdot,\cdot\}$ is Poisson bracket. On the other hand, one also has $dq_2/dt=\{q_2,\underline{H}_l\}=g\underline{L}_{z_1}$. Integrating, one thus obtains \begin{equation}
q_2(t)=q_2(0)+g\underline{L}_{z_1}t. 
\label{classical pointer}
\end{equation}
Hence, one can infer the value of $\underline{L}_{z_1}$ of the system (first particle) prior to the measurement from the initial and final positions of the apparatus (second particle). 

Let us now consider an ensemble of identically prepared angular momentum measurement and investigate the modification imposed by Eq. (\ref{fundamental equation}) to the classical dynamical equations that govern the ensemble of trajectories. To do this, first, given the classical Hamiltonian of Eq. (\ref{classical Hamiltonian angular momentum}), the Hamilton-Jacobi equation of (\ref{H-J equation}) becomes
\begin{equation}
\partial_t\underline{S}+g\big(x_1\partial_{y_1}\underline{S}-y_1\partial_{x_1}\underline{S}\big)\partial_{q_2}\underline{S}=0. 
\label{H-J equation angular momentum}
\end{equation}
On the other hand, substituting Eq. (\ref{classical Hamiltonian angular momentum}) into Eq. (\ref{classical velocity field}), the classical velocity field is given by 
\begin{eqnarray}
{\underline{v}}_{x_1}=-gy_1\partial_{q_2}\underline{S},\hspace{2mm}{\underline{v}}_{y_1}=gx_1\partial_{q_2}\underline{S},\hspace{2mm}{\underline{v}}_{z_1}=0,\nonumber\\
{\underline{v}}_2=g\big(x_1\partial_{y_1}\underline{S}-y_1\partial_{x_1}\underline{S}\big). \hspace{15mm}
\label{classical velocity angular momentum}
\end{eqnarray} 
The continuity equation of (\ref{continuity equation}) then becomes
\begin{eqnarray}
\partial_t\underline{\rho}-gy_1\partial_{x_1}(\underline{\rho}\partial_{q_2}\underline{S})+gx_1\partial_{y_1}(\underline{\rho}\partial_{q_2}\underline{S})\nonumber\\
+gx_1\partial_{q_2}(\underline{\rho}\partial_{y_1}\underline{S})-gy_1\partial_{q_2}(\underline{\rho}\partial_{x_1}\underline{S})=0. 
\label{continuity equation angular momentum}
\end{eqnarray}

Next, from Eq. (\ref{classical velocity angular momentum}) and the definition of $f$ given by Eq. (\ref{classical velocity field}), one has
\begin{eqnarray}
f_{x_1}(\underline{S})=-gy_1\partial_{q_2}\underline{S},\hspace{2mm}f_{y_1}=gx_1\partial_{q_2}\underline{S},\hspace{2mm}f_{z_1}(\underline{S})=0,\nonumber\\
f_2(\underline{S})=g\big(x_1\partial_{y_1}\underline{S}-y_1\partial_{x_1}\underline{S}\big), \hspace{15mm}
\label{actual velocity field angular momentum}
\end{eqnarray}
so that $\partial_q\cdot f(S)=2g(x_1\partial_{q_2}\partial_{y_1}S-y_1\partial_{q_2}\partial_{x_1}S)$. Substituting this into Eq. (\ref{fundamental equation}), one then obtains 
\begin{eqnarray}
\underline{\rho}\mapsto\Omega,\hspace{33mm}\nonumber\\
\partial_{x_1}\underline{S}\mapsto\partial_{x_1}S+\frac{\lambda}{2}\frac{\partial_{x_1}\Omega}{\Omega},\hspace{20mm}\nonumber\\
\partial_{y_1}\underline{S}\mapsto\partial_{y_1}S+\frac{\lambda}{2}\frac{\partial_{y_1}\Omega}{\Omega},\hspace{20mm}\nonumber\\
\partial_{q_2}\underline{S}\mapsto\partial_{q_2}S+\frac{\lambda}{2}\frac{\partial_{q_2}\Omega}{\Omega},\hspace{20mm}\nonumber\\
\partial_{t}\underline{S}\mapsto\partial_{t}S+\frac{\lambda}{2}\frac{\partial_{t}\Omega}{\Omega}+ g\lambda(x_1\partial_{y_1}\partial_{q_2}S-y_1\partial_{x_1}\partial_{q_2}S).\hspace{0mm}
\label{fundamental equation angular momentum}
\end{eqnarray}

Let us proceed to see how the above set of equations modify Eqs. (\ref{H-J equation angular momentum}) and (\ref{continuity equation angular momentum}). Imposing the first four equations of (\ref{fundamental equation angular momentum}) into Eq. (\ref{continuity equation angular momentum}) one obtains, after a simple calculation
\begin{eqnarray}
\partial_t\Omega-gy_1\partial_{x_1}(\Omega\partial_{q_2}S)+gx_1\partial_{y_1}(\Omega\partial_{q_2}S)+gx_1\partial_{q_2}(\Omega\partial_{y_1}S)\nonumber\\
-gy_1\partial_{q_2}(\Omega\partial_{x_1}S)-g\lambda(y_1\partial_{x_1}\partial_{q_2}\Omega-x_1\partial_{y_1}\partial_{q_2}\Omega)=0.\nonumber\\
\label{FPE angular momentum}
\end{eqnarray}
On the other hand, imposing the last four equations of (\ref{fundamental equation angular momentum}) into Eq. (\ref{H-J equation angular momentum}), one has, after an arrangement
\begin{eqnarray}
\partial_tS+g\big(x_1\partial_{y_1}S-y_1\partial_{x_1}S\big)\partial_{q_2}S-g\lambda^2\Big(x_1\frac{\partial_{y_1}\partial_{q_2}R}{R}\nonumber\\-y_1\frac{\partial_{x_1}\partial_{q_2}R}{R}\Big)+\frac{\lambda}{2\Omega}\Big(\partial_t\Omega-gy_1\partial_{x_1}(\Omega\partial_{q_2}S)\nonumber\\
+gx_1\partial_{y_1}(\Omega\partial_{q_2}S)+gx_1\partial_{q_2}(\Omega\partial_{y_1}S)-gy_1\partial_{q_2}(\Omega\partial_{x_1}S)\nonumber\\
-g\lambda(y_1\partial_{x_1}\partial_{q_2}\Omega-x_1\partial_{y_1}\partial_{q_2}\Omega)\Big)=0,
\label{ccc}
\end{eqnarray}
where we have defined a real-valued function $R\doteq\sqrt{\Omega}$ and used the following identity $(1/4)(\partial_{q_i}\Omega\partial_{q_j}\Omega/\Omega^2)=(1/2)(\partial_{q_i}\partial_{q_j}\Omega/\Omega)-(\partial_{q_i}\partial_{q_j}R/R)$. Substituting Eq. (\ref{FPE angular momentum}) into Eq. (\ref{ccc}), the last term in the bracket vanishes to give 
\begin{eqnarray}
\partial_tS+g\big(x_1\partial_{y_1}S-y_1\partial_{x_1}S\big)\partial_{q_2}S\hspace{20mm}\nonumber\\
-g\lambda^2\Big(x_1\frac{\partial_{y_1}\partial_{q_2}R}{R}-y_1\frac{\partial_{x_1}\partial_{q_2}R}{R}\Big)=0.
\label{HJM angular momentum}
\end{eqnarray}
The dynamics of ensemble of trajectories is then determined by pair of coupled Eqs. (\ref{FPE angular momentum}) and (\ref{HJM angular momentum}) which depend on the random hidden variable $\lambda$.   

Now let us assume that $\Omega$ satisfies the following condition: 
\begin{equation}
\Omega(q,\lambda;t)=\Omega(q,-\lambda;t),
\end{equation}
so that $P(\lambda)=\int dq\Omega(q,\lambda;t)=P(-\lambda)$ as required by Eq. (\ref{sub-quantum condition}). In this case, $S(q,\lambda;t)$ and $S(q,-\lambda;t)$ satisfy the same differential equation of (\ref{HJM angular momentum}), namely the last term on the left hand side is insensitive to the sign of $\lambda$. Hence, assuming that initially $S(q,\lambda;0)=S(q,-\lambda;0)$, one obtains 
\begin{equation}
S(q,\lambda;t)=S(q,-\lambda;t). 
\label{quantum phase symmetry}
\end{equation}
This can be used to eliminate the last term on the left hand side of Eq. (\ref{FPE angular momentum}). That is, taking the case when $\lambda$ is positive add to it the case when $\lambda$ is negative and divided by two one gets
\begin{eqnarray}
\partial_t\Omega-gy_1\partial_{x_1}(\Omega\partial_{q_2}S)+gx_1\partial_{y_1}(\Omega\partial_{q_2}S)+gx_1\partial_{q_2}(\Omega\partial_{y_1}S)\nonumber\\
-gy_1\partial_{q_2}(\Omega\partial_{x_1}S)=0.\nonumber\\
\label{quantum CE angular momentum}
\end{eqnarray}
 
Recalling that $\lambda\neq 0$ as required by Eq. (\ref{sub-quantum condition}), let us further define the following complex-valued (wave) function:
\begin{equation}
\Psi(q,\lambda;t)\doteq\sqrt{\Omega}\exp\Big(i\frac{S}{|\lambda|}\Big)=R\exp\Big(i\frac{S}{|\lambda|}\Big),  
\label{general wave function}
\end{equation}
where we have notationally omit the dependence of $\Psi$ on $\hbar$ and $\sigma$. From Eq. (\ref{marginal probabilities}), the probability density for the position of the particles is thus 
\begin{equation}
\rho(q;t)=\int d\lambda|\Psi|^2. 
\label{modified Born's statistical interpretation}
\end{equation}
Equations (\ref{HJM angular momentum}) and (\ref{quantum CE angular momentum}) can then be rewritten into the following generalized Schr\"odinger equation:
\begin{equation} 
i|\lambda|\partial_t\Psi=-g\lambda^2\big(x_1\partial_{y_1}-y_1\partial_{x_1}\big)\partial_{q_2}\Psi=g\frac{\lambda^2}{\hbar^2}\hat{L}_{z_1}\hat{p}_2\Psi,
\label{generalized Schroedinger equation angular momentum}
\end{equation}
where $\hat{L}_{z_1}\doteq -i\hbar(x_1\partial_{y_1}-y_1\partial_{x_1})$ and $\hat{p}_2\doteq-i\hbar\partial_{q_2}$ are the quantum mechanical $z-$angular momentum and linear momentum operators pertaining to the wave functions of the first and second particle, respectively, and we have assumed that the spatiotemporal fluctuations of $\lambda$ is ignorable as compared to that of $S$. 

Let us consider a specific case where $\Omega$ is separable $\Omega(q,\lambda;t,\hbar,\sigma)=\rho(q;t,\hbar,\sigma)P(\lambda;\hbar,\sigma)$ and the distribution of $\lambda$ is given by Eq. (\ref{quantum condition}) as assumed in Ref. \cite{AgungDQM1}. Further, let us define a new complex-valued function
\begin{equation}
\Psi_Q(q;t)\doteq\sqrt{\rho(q;t)}e^{\frac{i}{\hbar}S_Q(q;t)}, 
\end{equation}
where $S_Q(q;t)\doteq S(q,\pm\hbar;t)$. Then, Eqs. (\ref{generalized Schroedinger equation angular momentum})  reduces into the Schr\"odinger equation
\begin{equation}
i\hbar\partial_t\Psi_Q=\hat{H}_l\Psi_Q,
\label{Schroedinger equation angular momentum}
\end{equation}
with quantum Hamiltonian $\hat{H}_l$ 
\begin{equation}
{\hat H}_l\doteq g{\hat L}_{z_1}{\hat p}_2.
\label{quantum Hamiltonian angular momentum}
\end{equation} 
Equations (\ref{Schroedinger equation angular momentum}) and (\ref{quantum Hamiltonian angular momentum}) are the model employed by von Neumann to discuss quantum measurement \cite{von Neumann book}. The above result can be extended to the measurement of angular momentum along the $x-$ and $y-$ directions by cyclic permutation of $(x,y,z)$. In this case, $\hat{L}_{z_1}$ in Eq. (\ref{quantum Hamiltonian angular momentum}) is replaced by $\hat{L}_{x_1}$ and $\hat{L}_{y_1}$, the quantum mechanical angular momentum operators along the $x-$ and $y-$ directions, respectively. We have thus reproduced the results of canonical quantization as a specific case of our hidden variable model. We have also shown in Ref. \cite{AgungDQM1} that unlike canonical quantization, the above method of quantization is free from operator ordering ambiguity. 

\section{Possible corrections to quantum mechanical predictions}

In this section we shall go beyond quantum mechanics by assuming that the distribution of the hidden variable $\lambda$ satisfies Eq. (\ref{sub-quantum condition}) rather than Eq. (\ref{quantum condition}). We thus have to start from the generalized Schr\"odinger equation of (\ref{generalized Schroedinger equation angular momentum}). Various possible corrections to the prediction of quantum mechanics will be given. In general, the prediction of standard quantum mechanics will be argued to be reliable only for sufficiently low quantum number.  

\subsection{Hidden random variable induced broadening of spectral line\label{inherent broadening}} 

Let us discuss measurement of angular momentum in ensemble of identically prepared system so that the initial wave function of the system (first particle) $\psi(q_1)$ is given by one of the eigenfunction of the angular momentum operator $\psi(q_1)=\phi_l(q_1)$, ${\hat L}_{z_1}\phi_l=l\phi_l$, where $l$ is the eigenvalue. Further, let us denote the initial wave function of the apparatus (second particle) by $\varphi_0(q_2)$, assumed to be sufficiently localized. The total initial wave function of the system-apparatus is thus given by 
\begin{equation}
\Psi(q;0)=\phi_l(q_1)\varphi_0(q_2). 
\label{initial wave function angular momentum}
\end{equation}
We have thus made an idealization that the initial wave function is independent of $\lambda$. Recall that in this case, according to the standard quantum mechanics, each single measurement event will give outcome $l$ with certainty (probability one). This is one of the postulate of quantum mechanics. 

Let us solve Eq. (\ref{generalized Schroedinger equation angular momentum}) with the initial condition given by Eq. (\ref{initial wave function angular momentum}). To do this, let us assume that after interval time-span $t$ of measurement-interaction, the wave function can be written as 
\begin{equation}
\Psi(q,\lambda;t)=\phi_l(q_1)\varphi(q_2,\lambda;t). 
\label{anzat}
\end{equation}
Inserting this into Eq. (\ref{generalized Schroedinger equation angular momentum}) and keeping in mind that ${\hat L}_{z_1}\phi_l=l\phi_l$, one has 
\begin{equation}
\partial_t\varphi+gl'\partial_{q_2}\varphi=0, 
\label{von Neumann magic}
\end{equation}
where $l'$ depends on $\lambda$ as 
\begin{equation}
l'(\lambda)=\frac{|\lambda|}{\hbar}l. 
\label{actual angular momentum}
\end{equation}

Equation (\ref{von Neumann magic}) can then be directly integrated with the initial condition $\varphi(q_2,\lambda;0)=\varphi_0(q_2)$ to give
\begin{equation}
\varphi(q_2,\lambda;t)=\varphi_0(q_2-gl't). 
\label{broadening 0}
\end{equation}
Inserting this back into Eq. (\ref{anzat}), one has
\begin{equation}
\Psi(q,\lambda;t)=\phi_l(q_1)\varphi_0(q_2-g|\lambda|lt/\hbar). 
\end{equation}
Hence, in each single measurement event, the wave function of the apparatus becomes correlated to the initial state of the system and is shifted an amount of $gl'(\lambda)t$. This means that at the end of each single measurement event, the initial position of the second particle (the apparatus pointer) is shifted uniformly as 
\begin{equation}
q_2(t,\lambda)=q_2(0)+gl'(\lambda)t.
\label{apparatus pointer}
\end{equation} 

Now let us interpret the above formalism in similar way as with classical measurement. As discussed in the previous section, in the latter case, after time-span of measurement-interaction $t$, the position of the apparatus-particle is shifted as $q_2(t)=q_2(0)+g\underline{L}_{z_1}t$. From this, one infers the result of measurement to be given by $\underline{L}_{z_1}$. Similarly, it is natural to interpret Eq. (\ref{apparatus pointer}) that the outcome of each single measurement event is given by $l'(\lambda)=|\lambda|l/\hbar$ of Eq. (\ref{actual angular momentum}). Here we have applied the result shown in Ref. \cite{AgungDQM1} that it is possible to probe the pre-existing value of the initial and final positions of the apparatus particle \cite{position measurement}. Hence, instead of obtaining a sharp value $l$ as postulated by the standard quantum mechanics, one obtains a random value $l'(\lambda)$ which depends on the value of the hidden variable $\lambda$. One can also see that when the distribution of $\lambda$ is given by Eq. (\ref{quantum condition}) so that $\lambda=\pm\hbar$, then the randomness of the outcome of single measurement disappears and one regains the prediction of quantum mechanics: $l'(\pm\hbar)=l$ with probability one. For general distribution of $\lambda$ satisfying Eq. (\ref{sub-quantum condition}), we have thus a random correction to the prediction of quantum mechanics: even when the initial wave function of the system is given by one of the eigenfunction of the angular momentum operator, the result of each single measurement will still be random with statistical properties determined by the distribution of $\lambda$. 

Hence, given the value of $l$, the probability density to get $l'$ is  
\begin{eqnarray}
P(l'|l)=\frac{\hbar}{|l|}\Big(f_+(\lambda;\hbar,\sigma)|_{\lambda=\frac{l'}{l}\hbar}+f_-(\lambda;\hbar,\sigma)|_{\lambda=-\frac{l'}{l}\hbar}\Big)\nonumber\\
=2\frac{\hbar}{|l|}f_+(\lambda;\hbar,\sigma)|_{\lambda=\frac{l'}{l}\hbar},\hspace{10mm} 
\label{conditional probability: eigen state}
\end{eqnarray}
where $f_+(\lambda;\hbar,\sigma)$ and $f_-(\lambda;\hbar,\sigma)$ are part of $P(\lambda;\hbar,\sigma)$ defined on the positive and negative axis of $\lambda$ respectively, and in the second equality we have used the assumption that $f_+(\lambda;\hbar,\sigma)=f_-(-\lambda;\hbar,\sigma)$ of Eq. (\ref{sub-quantum condition}). In the limit $\sigma\rightarrow 0$ one has $\lim_{\sigma\rightarrow 0}f_+(\lambda;\hbar,\sigma)=(1/2)\delta(\lambda-\hbar)$ by Eq. (\ref{quantum condition}) so that we reproduce the prediction of quantum mechanics
\begin{equation}
\lim_{\sigma\rightarrow 0}P(l'|l)=\frac{\hbar}{|l|}\delta\Big(\frac{\hbar}{l}(l'-l)\Big)=\delta(l'-l),    
\label{quantum mechanical condition probability: eigen state}
\end{equation}
that is, in each single measurement event, one always obtains $l'=l$, as expected. 

Let us proceed to discuss the statistical properties of $l'$ in term of the statistical properties of the hidden variable $\lambda$, $P(\lambda;\hbar,\sigma)$. Recall that $P(\lambda;\hbar,\sigma)$ must satisfy Eq. (\ref{sub-quantum condition}). There are then infinitely many $P(\lambda;\hbar,\sigma)$ fulfilling this requirement. Let us give a general method to construct such probability density function. First, since $P(\lambda;\hbar,\sigma)=P(-\lambda;\hbar,\sigma)$ then it is sufficient to fix the form of $P(\lambda;\hbar,\sigma)$ on the half line $\lambda> 0$. Let us then pick up a non-negative function denoted by $P_{+}(\lambda;\hbar,\sigma)$ which is defined on $\lambda>0$. Further, let us assume that $P_+(\lambda;\hbar,\sigma)$ is normalizable, $\int_0^{\infty} d\lambda P_{+}(\lambda;\hbar,\sigma)=1$, and possessing the following limiting property:
\begin{equation}
\lim_{\sigma\rightarrow 0}P_{+}(\lambda;\hbar,\sigma)=\delta(\lambda-\hbar).  
\label{sub-quantum condition 2}
\end{equation}  
$\sigma$ thus measures the width of $P_+(\lambda;\hbar,\sigma)$. The desired probability density for the hidden random variable $\lambda$ can then be constructed as 
\begin{equation}
P(\lambda;\hbar,\sigma)=\frac{1}{2}P_{+}(\lambda;\hbar,\sigma)U(\lambda)+\frac{1}{2}P_{+}(-\lambda;\hbar,\sigma)U(-\lambda), 
\label{general hidden random variable}
\end{equation}
where $U(\lambda)$ is the Heaviside step-function, namely $U(\lambda)=1$ for $\lambda\ge 0$ and $U(\lambda)=0$ for $\lambda<0$. It is then evident that the so-constructed $P(\lambda;\hbar,\sigma)$ possesses the required symmetry property $P(\lambda;\hbar,\sigma)=P(-\lambda;\hbar,\sigma)$. Moreover, in the formal limit $\sigma\rightarrow 0$, one obtains, by the virtue of Eq. (\ref{sub-quantum condition 2})
\begin{equation}
\lim_{\sigma\rightarrow 0}P(\lambda;\hbar,\sigma)=\frac{1}{2}\delta(\lambda-\hbar)+\frac{1}{2}\delta(\lambda+\hbar)=P_Q(\lambda;\hbar),
\end{equation} 
as required by Eq. (\ref{sub-quantum condition}). The prediction of quantum mechanics is thus regained in the limit of vanishing $\sigma$ which is equal to the vanishing of the width of $P_+(\lambda;\hbar,\sigma)$.

The mean and variance of the fluctuation of $l'$ conditioned on the value of $l$ (quantum number) can then be expressed as follows. First, given $l$, the mean of $l'$ is 
\begin{equation}
M_1[P(l'|l)]=\frac{l}{\hbar}\int_{-\infty}^{\infty}d\lambda |\lambda|P(\lambda;\hbar,\sigma)=\frac{l}{\hbar}M_1[P_{+}(\lambda;\hbar,\sigma)]. 
\label{mean value for eigenstate}
\end{equation}
Notice that in the limit $\sigma\rightarrow 0$, one has $M_1[P(\lambda;\hbar,\sigma\rightarrow 0)]=\hbar$  by Eq. (\ref{sub-quantum condition 2}), so that one regains the prediction of quantum mechanics $M_1[P(l'|l)]\rightarrow l$. In general, for non-vanishing $\sigma$, however, $M_1[P_+(\lambda;\hbar,\sigma)]\neq\hbar$ and is independent of $l$ so that there is a correction to the prediction of quantum mechanics which is proportional to $l$ (the value predicted by quantum mechanics). Similarly, the second moment is given by  
\begin{equation}
M_2[P(l'|l)]=\frac{l^2}{\hbar^2}\int_{0}^{\infty}d\lambda\lambda^2P_+(\lambda;\hbar,\sigma).
\label{second moment for eigen state}
\end{equation}
The variance of $l'$ given the value of $l$ is thus 
\begin{equation}
\mbox{Var}[P(l'|l)]=\frac{l^2}{\hbar^2}\mbox{Var}[P_{+}(\lambda;\hbar,\sigma)].
\label{variance for eigenstate} 
\end{equation}
Again in the limit $\sigma\rightarrow 0$, one regains the prediction of quantum mechanics $\mbox{Var}[P(l'|l)]\rightarrow 0$, by Eq. (\ref{sub-quantum condition 2}). Hence, in general for non-vanishing $\sigma$, there is a finite broadening of the spectral line given by the width of $P_+(\lambda;\hbar,\sigma)$ and is proportional to $l$. 

Let us take a concrete statistical model by assuming the following form of $P_{+}(\lambda;\hbar,\sigma)$: 
\begin{equation}
P_{+}(\lambda;\hbar,\sigma)=\frac{1}{\lambda\sqrt{2\pi\sigma^2}}\exp\Big\{-\frac{(\ln\lambda-\ln\hbar)^2}{2\sigma^2}\Big\},\hspace{2mm}\lambda> 0.
\label{log-normal pdf}
\end{equation}
It is the log-normal distribution with location parameter $\ln\hbar$, scale parameter $\sigma$ and thus mode (the position of its maximum) $\lambda_M=\hbar\exp(-\sigma^2)$;  $x\doteq\ln\lambda$ is normally distributed with mean $\ln\hbar$ and width $\sigma$ \cite{log-normal pdf}. Hence, in the limit of $\sigma\rightarrow 0$ one indeed has $\lim_{\sigma\rightarrow 0}P_{+}(\lambda;\hbar,\sigma)=\delta(\lambda-\hbar)$, as required by Eq. (\ref{sub-quantum condition 2}). The mean, second moment and variance are given by 
\begin{eqnarray}
M_1[P_+(\lambda;\hbar,\sigma)]=\hbar e^{\sigma^2/2},\hspace{2mm}M_2[P_+(\lambda;\hbar,\sigma)]=\hbar^2e^{2\sigma^2}\nonumber\\
\mbox{Var}[P_+(\lambda;\hbar,\sigma)]=\hbar^2e^{\sigma^2}(e^{\sigma^2}-1). \hspace{10mm}
\end{eqnarray}

In this case, the profile of the broadening of the quantum mechanical spectral line with quantum number $l$ can be obtained by inserting Eq. (\ref{log-normal pdf}) into Eq. (\ref{conditional probability: eigen state}), noticing $f_+(\lambda;\hbar,\sigma)=P_+(\lambda;\hbar,\sigma)/2$, to give 
\begin{equation}
P(l'|l)=\frac{1}{|l'|\sqrt{2\pi}\sigma}\exp\Big\{-\frac{(\ln l'/l)^2}{2\sigma^2}\Big\}, 
\label{profile of broadening}
\end{equation}
where $l'/l>0$. The profile of the broadening is given by the log-normal function and thus not symmetric. Next, the conditional average of $l'$ given in Eq. (\ref{mean value for eigenstate}) is
\begin{equation}
M_1[P(l'|l)]=le^{\sigma^2/2}\approx l+l\frac{\sigma^2}{2}+O(\sigma^4).
\label{average angular momentum-eigenfunction}
\end{equation} 
Hence, the absolute value of the average of $l'$ is always larger than the prediction of quantum mechanics. The correction to the prediction of quantum mechanics is proportional to the latter and also to the value of $\sigma^2$. Further, the second moment of $l'$ conditioned on the value of $l$ is given by $M_2[P(l'|l)]=l^2e^{2\sigma^2}$. The variance of $l'$ given the value of $l$ is thus  
\begin{equation}
\mbox{Var}[P(l'|l)]=l^2e^{\sigma^2}(e^{\sigma^2}-1)\approx l^2\sigma^2+O(\sigma^4).  
\end{equation}
In the quantum limit where $\sigma\rightarrow 0$ we regain the prediction of quantum mechanics: $M_1[P(l'|l)]\rightarrow l$, $M_2[P(l'|l)]\rightarrow l^2$ and $\mbox{Var}[P(l'|l)]\rightarrow 0$. 

The above results suggest that the prediction of quantum mechanics is reliable only for sufficiently low quantum number $|l|$. Namely, the deviation from the prediction of quantum mechanics grows as $|l|$ increases. In particular for a symmetric log-normal model with a given $\sigma$, the prediction of quantum mechanics is ambiguous for $|l|$ satisfying $|l|\sigma^2/2\approx\Delta l$, where $\Delta l$ is the quantum mechanical spectral  spacing. In general statistical model, for sufficiently large value of $|l|$, $\mbox{Var}[P(l'|l)]\sim (\Delta l)^2$ so that the quantum discreteness is smoothed out \cite{high quantum number vs classical limit}. 

\subsection{Modified Born's statistical rule}

Now let us consider the general case when the initial wave function of the system $\psi(q_1)$ is not necessarily the eigenfunction of the angular momentum operator. To do this, first, notice that Eq. (\ref{generalized Schroedinger equation angular momentum}) is linear with respect to $\Psi$. Hence, since $\phi_l(q_1)\varphi_0(q_2-gl|\lambda|t/\hbar)$ satisfies Eq. (\ref{generalized Schroedinger equation angular momentum}) as shown in the previous subsection, their linear superposition over all possible values of $l$
\begin{equation}
\Psi(q,\lambda;t)=\sum_lc_l\phi_l(q_1)\varphi_0(q_2-gl|\lambda|t/[\hbar), 
\label{general solution}
\end{equation}
also satisfies Eq. (\ref{generalized Schroedinger equation angular momentum}). Here $\{c_l\}$ is a set of complex numbers to be determined as follows. Putting $t=0$, one obtains 
\begin{equation}
\Psi(q;0)=\Big(\sum_lc_l\phi_l(q_1)\Big)\varphi_0(q_2),
\label{general initial state}
\end{equation}
which is separable and independent of $\lambda$. Hence, the initial wave function of the system alone (the first particle) is given by 
\begin{equation}
\psi(q_1)=\sum_lc_l\phi_l(q_1). 
\end{equation}
This shows that $c_l$ is the coefficient of expansion of the initial wave function of the system $\psi(q_1)$ in term of the set of orthonormal eigenfunctions of the angular momentum operator $\{\phi_l\}$, $c_l=\langle \phi_l|\psi\rangle$. 

One can then see from Eqs. (\ref{general solution}) and (\ref{general initial state}) that at the end of the measurement-interaction, the wave function of the apparatus-particle separates into a series of packets, each is correlated to one of the eigenfunction of the angular momentum operator. Namely, for a given value of $\lambda$, the wave packet of the apparatus is shifted an amount of $gl'(\lambda)t=g|\lambda|lt/\hbar$. Hence, if $\lambda$ is fixed and $\varphi_0(q_2)$ is spatially localized, then for sufficiently large value of $g$ 
\begin{equation}
\varphi_l(q_2,\lambda;t)\doteq\varphi_0(q_2-gl|\lambda|t/\hbar),
\label{shifted aparatus wave function}
\end{equation}
does not overlap with each other for different values of $l$, and each is correlated to a distinct eigenfunction of angular momentum operator, $\phi_l(q_1)$. 

Let us now denote the probability density that the apparatus-particle enters the support of the wave packet $\varphi_l(q_2,\lambda;t)$ as $P_{\varphi_l}$. Then, the probability density to get the value $l'$ is given by
\begin{equation}
P(l')=\sum_lP(l'|l)P_{\varphi_l},
\label{general measurement formula}
\end{equation} 
where $P(l'|l)$ is the probability density to get $l'$ provided that the apparatus-particle is inside the support of  $\varphi_l$ which is discussed in the previous subsection and is given by Eq. (\ref{conditional probability: eigen state}). 

It thus remains to calculate $P_{\varphi_l}$. To do this, first, since for sufficiently large value of $g$, $\varphi_l(q_2,\lambda;t)$ in Eq. (\ref{general solution}) does not overlap for different values of $l$, then the joint-probability density that the first particle (system) is at $q_1$ and the second particle (apparatus) is at $q_2$ for a given value of $\lambda$ is decomposed into  
\begin{equation}
\Omega(q,\lambda;t)=|\Psi(q,\lambda;t)|^2\approx\sum_l|c_l|^2|\phi_l(q_1)|^2|\varphi_l(q_2,\lambda;t)|^2. 
\label{decomposition-approximation}
\end{equation}
Let us note that when $|\lambda|$ is very small one needs a large value of $g$ to separate $\varphi_l$ for different values of $l$ otherwise the above decomposition is not valid. Nevertheless, since $P(\lambda;\hbar,\sigma)$ is very small in the regime where $|\lambda|\ll \hbar$, one can argue that its contribution is ignorable. 

From Eq. (\ref{decomposition-approximation}), one can see that the joint-probability density that the first particle has coordinate $q_1$ and the second particle has coordinate $q_2$ inside the support of $\varphi_l$ for a fixed value of $\lambda$ is given by  
\begin{equation}
|c_l|^2|\phi_l(q_1)|^2|\varphi_l(q_2,\lambda;t)|^2. 
\label{joint-probability position-labelling}
\end{equation}
The probability density that the second particle is inside the support of the wave packet $\varphi_l$ regardless of the position of the first and second particles and the value of $\lambda$ is thus 
\begin{equation}
P_{\varphi_l}=\int dq_1dq_2d\lambda|c_l|^2|\phi_l(q_1)|^2|\varphi_l(q_2,\lambda;t)|^2=|c_l|^2,
\label{Born rule}
\end{equation}
which is just the Born's statistical rule. 

Finally, inserting Eq. (\ref{Born rule}) into Eq. (\ref{general measurement formula}), the probability density to get $l'$ can be calculated as
\begin{equation}
P(l')=\sum_lP(l'|l)|c_l|^2.
\label{modified Born's rule}
\end{equation}
We have thus a modified Born's statistical rule. Since in the limit $\sigma\rightarrow 0$, $P(\lambda;\hbar,\sigma)$ reduces into $P_Q(\lambda;\hbar)$ given by Eq. (\ref{quantum condition}) so that one has $P(l'|l)\rightarrow\delta(l'-l)$ of Eq. (\ref{quantum mechanical condition probability: eigen state}), then Eq. (\ref{modified Born's rule}) reduces into 
\begin{equation}
\lim_{\sigma\rightarrow 0}P(l')=\sum_l|c_l|^2\delta(l'-l),
\end{equation}
as postulated by quantum mechanics.   

Next, using Eqs. (\ref{modified Born's rule}) and (\ref{mean value for eigenstate}), the average of $l'$ can be calculated to give 
\begin{equation}
M_1[P(l')]=\int dl' l' P(l')=\frac{M_1[P_+(\lambda;\hbar,\sigma)]}{\hbar}M_Q. 
\label{correction to average}
\end{equation}
where $M_Q=\sum_l l|c_l|^2$ is the quantum mechanical prediction for the average value of $l$. Hence, there is a correction which depends on the value of $\sigma$. If the initial wave function of the system is $\phi_l$, one has $M_Q=l$ so that one regains Eq. (\ref{mean value for eigenstate}). For the case where $P_+(\lambda;\hbar,\sigma)$ is log-normal function given by Eq. (\ref{log-normal pdf}), then $M_1[P_+(\lambda;\hbar,\sigma)]=\hbar\exp(\sigma^2/2)$ so that one has 
\begin{equation}
M_1[P(l')]=M_Q e^{\sigma^2/2}\approx M_Q(1+\sigma^2/2)+O(\sigma^4).  
\label{correction to average log-normal}
\end{equation} 
Further, using Eq. (\ref{second moment for eigen state}), the variance can be expressed as, after an arrangement, 
\begin{equation}
\mbox{Var}[P(l')]=\mbox{VarQ}\frac{M_2[P_+(\lambda;\hbar,\sigma)]}{\hbar^2}+M_Q^2\frac{\mbox{Var}[P_+(\lambda;\hbar,\sigma)]}{\hbar^2}, 
\label{correction to variance}
\end{equation}
where $\mbox{VarQ}=\sum_l(l-M_Q)^2|c_l|^2$ is the variance predicted by quantum mechanics. Again, if the initial wave function of the system is $\phi_l$, then $\mbox{VarQ}=0$ and $M_Q=l$ so that one regains Eq. (\ref{variance for eigenstate}) as expected. Assuming that $P_+(\lambda;\hbar,\sigma)$ takes the form of log-normal function of Eq. (\ref{log-normal pdf}), one has 
\begin{eqnarray}
\mbox{Var}[P(l')]=\mbox{VarQ}e^{2\sigma^2}+M_Q^2(e^{\sigma^2}-1)e^{\sigma^2}\nonumber\\
\approx\mbox{VarQ}(1+2\sigma^2)+M_Q^2\sigma^2+O(\sigma^4). 
\label{correction to variance log-normal}
\end{eqnarray}
From Eqs. (\ref{sub-quantum condition 2}), (\ref{correction to average}) and (\ref{correction to variance}), one can see that the prediction of quantum mechanics is regained in the ``quantum limit'' $\sigma\rightarrow 0$. 

All the above corrections also apply to measurement of other physical quantities. In particular, we have shown in Ref. \cite{AgungDQM1} that the measurement of momentum and position can also be treated in the same way as the measurement of angular momentum discussed in the present paper, so that the corrections to the prediction of quantum mechanics obtained above also apply to the measurement of position and momentum. Since in general for non-vanishing $\sigma$ one has $M_2[P_+(\lambda;\hbar,\sigma)]\ge\hbar^2$, one can then conclude that the variance of the measurement results predicted by the present hidden variable model is in general larger than the prediction of the quantum mechanics. Hence, for the case of measurement of position and momentum of identically prepared system, we will have an inequality which is in general stronger than the Heisenberg uncertainty relation. 

Further, it is also evident from the above discussion that there are two sources of randomness of the outcomes of a single measurement event. The first kind of randomness comes from our ignorance of the initial position of the particles. As discussed above, this type of randomness determines the probability that the apparatus particle enters the support of the wave packet $\varphi_l=\varphi_0(q_2-gl|\lambda|t/\hbar)$ which is given by $P_{\varphi_l}=|c_l|^2$. This is the only type of source of randomness asserted by pilot-wave theory \cite{Bohm paper}. The second type of randomness comes from the fluctuations of $\lambda$ which determines the probability to get $l'$ provided that the apparatus particle is inside the support of $\varphi_l$, $P(l'|l)$. This latter kind of source of randomness of single measurement event thus is inherent in the dynamics.  

Let us further discuss the important role played by $\lambda$. In Ref. \cite{AgungDQM1}, we have shown that in the limit $\sigma\rightarrow 0$ our hidden variable model effectively reproduces the mathematical formalism of pilot-wave theory  \cite{Bohm paper}. In most literatures of pilot-wave theory, position is regarded as hidden variable. However, it is through the position of the particle that we experience the real world. Noticing this fact, Bell then proposed to regard the wave function, which is more hidden to us and is assumed to be physically real, as the hidden variable; and call the position as the dynamical beable \cite{Bell beable}. In this context, $\lambda$ in our statistical model allows us to omit the necessity to assume the wave function as physically real. In our model, the wave function is only an artificial calculational tool. Its status as hidden variable is replaced by $\lambda$. Moreover, in contrast to pilot-wave theory which is deterministic, $\lambda$ makes the model inherently stochastic. It is also the main ingredient of the quantization processes through which we get the Schr\"odinger equation with a unique physical interpretation. Hence, it is $\lambda$ that distinguishes ``quantum-ness'' from classical statistical mechanics.

\subsection{Stern-Gerlach experiment\label{Stern-Gerlach experiment}}

To give a concrete example and to show the robustness of the statistical model, let us now apply the statistical model developed in the previous subsections to investigate the Stern-Gerlach experiment of measurement of angular momentum. As will be clear, Stern-Gerlach type of measurement is different from the von Neumann model of measurement discussed in the previous subsections. 

First, let us discuss the description of ensemble of the Stern-Gerlach experiment in classical dynamics. Let us assume that we have a beam of neutral atom whose center of mass coordinate is denoted by $q_2=(x_2,y_2,z_2)$, each containing an electron with coordinate $q_1=(x_1,y_1,z_1)$. The interaction between the atom and the magnetic field of the Stern-Gerlach apparatus is thus mainly due to angular momentum of the electron. Let us assume that this interaction is impulsive so that one can neglect the free Hamiltonian of the system during the measurement-interaction. Further, let us assume that the magnetic field is non-vanishing only in $z-$direction $B=(0,0,B_z)$, $B_z=B'z_2$, where $B'$ is some constant, namely it is a monotonic function along the $z-$axis \cite{some complication}. In this case, the classical interaction-Hamiltonian can be approximated to be given by
\begin{equation} 
\underline{H}_{SG}\approx\frac{eB'}{2m_ec}z_2 \underline{L}_{z_1}=\frac{eB'}{2m_ec}z_2(x_1\underline{p}_{y_1}-y_1\underline{p}_{x_1}),
\label{classical Hamiltonian SG angular momentum}
\end{equation}
where $e$ is charge of electron, $m_e$ is its mass, $c$ is the velocity of light. We have also ignored the quantum mechanical spin degree of freedom. 

The Hamilton-Jacobi equation of (\ref{H-J equation}) thus reads
\begin{equation}
\partial_t\underline{S}+\zeta(z_2)(x_1\partial_{y_1}\underline{S}-y_1\partial_{x_1}\underline{S})=0,
\label{H-J equation SG angular momentum}
\end{equation}
where we have defined $\zeta(z_2)\doteq\frac{eB'}{2m_ec}z_2$. On the other hand, from the Hamilton equation of (\ref{classical velocity field}), one obtains the following classical velocity field
\begin{equation}
\underline{v}_{x_1}=-\zeta(z_2)y_1,\hspace{2mm}\underline{v}_{y_1}=\zeta(z_2)x_1,\hspace{2mm}\underline{v}_{z_1}=0,\hspace{2mm}\underline{v}_{z_2}=0. 
\label{classical velocity field SG angular momentum}
\end{equation}
The above set of equations give constraints to the dynamics. Using Eq. (\ref{classical velocity field SG angular momentum}), the continuity equation of (\ref{continuity equation}) becomes
\begin{equation}
\partial_t\underline{\rho}+\zeta(z_2)(x_1\partial_{y_1}\underline{\rho}-y_1\partial_{x_1}\underline{\rho})=0. 
\label{continuity equation SG angular momentum}
\end{equation}
Hence, the classical dynamics of ensemble of trajectories during interaction with Stern-Gerlach magnetic field is given by solving Eqs. (\ref{H-J equation SG angular momentum}) and (\ref{continuity equation SG angular momentum}) in terms of $\underline{S}(q,t)$ and $\underline{\rho}(q,t)$. 

Next, from Eq. (\ref{classical velocity field SG angular momentum}), $f$ defined in Eq. (\ref{classical velocity field}) is given by 
\begin{eqnarray}
f_{x_1}(\underline{S})=-\zeta(z_2)y_1,\hspace{2mm}f_{y_1}(\underline{S})=\zeta(z_2)x_1,\nonumber\\
f_{z_1}(\underline{S})=0,\hspace{2mm}f_{z_2}(\underline{S})=0, \hspace{10mm}
\label{magic SG angular momentum}
\end{eqnarray}
so that one has $\sum_i\partial_{q_i}f_i(S)=0$. Equation (\ref{fundamental equation}) then becomes 
\begin{eqnarray}
\underline{\rho}\mapsto\Omega,\hspace{23mm}\nonumber\\
\partial_{x_1}\underline{S}\mapsto\partial_{x_1}S+\frac{\lambda}{2}\frac{\partial_{x_1}\Omega}{\Omega},\hspace{10mm}\nonumber\\
\partial_{y_1}\underline{S}\mapsto\partial_{y_1}S+\frac{\lambda}{2}\frac{\partial_{y_1}\Omega}{\Omega},\hspace{10mm}\nonumber\\
\partial_t\underline{S}\mapsto\partial_tS+\frac{\lambda}{2}\frac{\partial_t\Omega}{\Omega}.\hspace{15mm}
\label{fundamental equation SG angular momentum} 
\end{eqnarray}

Let us investigate how the above set of equations modify Eqs. (\ref{H-J equation SG angular momentum}) and (\ref{continuity equation SG angular momentum}). First, since Eq. (\ref{continuity equation SG angular momentum}) does not contain $\underline{S}$, then imposing the first equation of (\ref{fundamental equation SG angular momentum}), one has 
\begin{equation}
\partial_t\Omega+\zeta(z_2)(x_1\partial_{y_1}\Omega-y_1\partial_{x_1}\Omega)=0. 
\label{continuity equation SG angular momentum: quantum}
\end{equation}
Further, imposing the last three equations of (\ref{fundamental equation SG angular momentum}) into Eq. (\ref{H-J equation SG angular momentum}) one has 
\begin{eqnarray}
\partial_tS+\zeta(z_2)(x_1\partial_{y_1}S-y_1\partial_{x_1}S)\hspace{20mm}\nonumber\\
+\frac{\lambda}{2\Omega}\big(\partial_t\Omega+\zeta(z_2)(x_1\partial_{y_1}\Omega-y_1\partial_{x_1}\Omega)\big)=0.
\end{eqnarray} 
Substituting Eq. (\ref{continuity equation SG angular momentum: quantum}), the above equation becomes
\begin{equation}
\partial_tS+\zeta(z_2)(x_1\partial_{y_1}S-y_1\partial_{x_1}S)=0.
\label{HJM equation SG angular momentum}
\end{equation}
We have thus pair of equations (\ref{continuity equation SG angular momentum: quantum}) and (\ref{HJM equation SG angular momentum}) which are similar to its classical counterpart of Eqs. (\ref{continuity equation SG angular momentum}) and (\ref{H-J equation SG angular momentum}).  

Finally, defining complex-valued function as 
\begin{equation}
\Psi_M\doteq R\exp(iS/\hbar),
\end{equation}
where $R\doteq\sqrt{\Omega}$, Eqs. (\ref{continuity equation SG angular momentum: quantum}) and (\ref{HJM equation SG angular momentum}) can be recast into the following Schr\"odinger equation:
\begin{equation}
i\hbar\partial_t\Psi_M=\zeta(z_2){\hat L}_{z_1}\Psi_M.
\label{Schroedinger equation SG angular momentum}
\end{equation} 

Now let us discuss the detail process of the Stern-Gerlach experiment. Let us first assume that the initial wave function of the total system prior to entering the Stern-Gerlach magnetic system is separable given by 
\begin{equation}
\Psi_M(q_1,q_2;0)=\phi_l(q_1)\varphi_0(q_2),
\end{equation} 
where again $\phi_l$ is the eigenfunction of quantum mechanical angular momentum operator ${\hat L}_{z_1}$ corresponding to eigenvalue $l$: ${\hat L}_{z_1}\phi_l=l\phi_l$. Let us assume that after spending time $t$ inside the magnetic system of the Stern-Gerlach apparatus, the wave function is still separable given by 
\begin{equation}
\Psi_M(q_1,q_2;t)=\phi_l(q_1)\varphi(q_2;t),
\label{wave function SG angular momentum}
\end{equation}
where $\varphi(q_2;0)=\varphi_0(q_2)$. Inserting Eq. (\ref{wave function SG angular momentum}) into the Schr\"odinger equation of (\ref{Schroedinger equation SG angular momentum}) one obtains
\begin{equation}
i\hbar\partial_t\varphi=\zeta(z_2)l\varphi,
\label{SG equation} 
\end{equation}
which can then be directly integrated to give 
\begin{equation}
\varphi(q_2;t)=\varphi_0(q_2)\exp(-i \mu ltz_2/\hbar),
\end{equation}
where $\mu\doteq eB'/(2m_ec)$. Inserting this back into Eq. (\ref{wave function SG angular momentum}), the total wave function at the exit of the Stern-Gerlach magnetic system at $t=T$ is thus  
\begin{eqnarray}
\Psi_M(q_1,q_2;T)=\phi_l(q_1)\varphi_0(q_2)e^{-\frac{i}{\hbar}\Delta_lz_2},
\label{superposition state SG angular momentum}
\end{eqnarray} 
where we have defined $\Delta_l\doteq \mu l T$. One can thus see that the atomic wave function gets a $l-$dependent phase with a wave vector along the $z-$direction given by $\Delta_l$. 

Now, after passing through the Stern-Gerlach magnet, let us assume that the atom is free. Thus the time evolution afterward is governed by a generalized Schr\"odinger equation 
\begin{equation}
i|\lambda|\partial_t\Psi(q,\lambda;t)=-\frac{\lambda^2}{2m_a}\partial_{z_2}^2\Psi(q,\lambda;t), 
\label{free atom Schroedinger}
\end{equation}
where $m_a$ is the mass of the atom and we have ignored the electronic free Hamiltonian and the irrelevant $x_2$ and $y_2$ part of the atomic wave function. The detail derivation of Eq. (\ref{free atom Schroedinger}) by imposing Eq. (\ref{fundamental equation}) to classical dynamics of ensemble with classical Hamiltonian $\underline{H}=\underline{p}_{z_2}^2/(2m_a)$ is given in \cite{AgungDQM1}. Equation (\ref{free atom Schroedinger}) has to be solved subject to the initial wave function at $t=0$ given by Eq. (\ref{superposition state SG angular momentum}): $\Psi(q;0)=\Psi_M(q;T)$. To do this, let us take a concrete model when the initial atomic wave function is Gaussian 
\begin{equation}
\varphi_0(z_2)\sim\exp\Big(-\frac{z_2^2}{4\sigma_0^2}\Big),
\end{equation}
up to normalization constant, where $\sigma_0$ is the width of the Gaussian. The Schr\"odinger equation of (\ref{free atom Schroedinger}) can then be solved exactly to give 
\begin{equation}
\Psi(q;t,\lambda)\sim\phi_l(q_1)e^{-\frac{(z_2-\frac{\Delta_l}{m_a}\frac{|\lambda|}{\hbar}t)^2}{4\sigma_t(\lambda)\sigma_0}-i\frac{\Delta_l}{\hbar}(z_2-\frac{1}{2}\frac{\Delta_l}{m_a}\frac{|\lambda|}{\hbar}t)},
\label{final wave function}
\end{equation} 
where $\sigma_t(\lambda)=\sigma_0(1+i|\lambda| t/2m_a\sigma_0^2)$. 

One can thus see from Eq. (\ref{final wave function}) that at time $t$ (measured just after the the particle leaving the magnetic field of the Stern-Gerlach apparatus), the initial atomic wave function is uniformly shifted by an amount 
\begin{equation}
O_l(t;\lambda)=\frac{\Delta_l}{m_a}\frac{|\lambda|}{\hbar}t=\frac{g_Ml|\lambda|t}{\hbar},\hspace{2mm}g_M\doteq\frac{\mu T}{m_a}. 
\end{equation}
We can thus admit that result of single measurement event is random given by $l'=|\lambda|l/\hbar$ reproducing the results obtained using the von Neumann model as discussed in the previous subsections. Hence, we can proceed as in the previous subsections to derive various corrections to the prediction of quantum mechanics by assuming a statistical distribution of $\lambda$. The above scheme can also be straightforwardly generalized to arbitrary initial electronic wave function $\psi(q_1)$. Hence, the precision of Stern-Gerlach experiment conducted so far can be regarded to give the upper bound for the assumed small yet finite value of $\sigma$. 

\section{Conclusion and discussion}

Following Ref. \cite{AgungDQM1}, we have modified the classical dynamics of ensemble of trajectories parameterized by hidden random variable to simulate the quantum fluctuations. The hidden variable $\lambda$ is  assumed to be non-vanishing, its probability density function satisfies a general symmetry condition of Eq. (\ref{sub-quantum condition}) so that it is unbiased, independent of time, and is characterized by the reduced Planck constant $\hbar$ and a real-valued dimensionless parameter $\sigma$. The statistical model is then applied to discuss the measurement of angular momentum in two different models, that of von Neumann and Stern-Gerlach, over an ensemble of identically prepared system. 

We showed that the prediction of quantum mechanics is regained in the formal limit $\sigma\rightarrow 0$, so that for non-vanishing $\sigma\ll 1$ it attains small corrections. First, we showed that there is a finite broadening on the spectral line which is purely due to the fluctuations of the hidden variable. Accordingly, the Born's statistical rule has to be slightly modified. In general, the correction to the prediction of quantum mechanics is larger for higher quantum number so that quantum mechanics is reliable for sufficiently low quantum number. These results thus allow precision tests of quantum mechanics against our hidden variable model. Let us mention that there are reports on precision test of quantum mechanics against possible nonlinear modification of Schr\"odinger equation \cite{precision test NLSE}. One of the interesting feature of our model, in this respect, is that possible deviations from the prediction of quantum mechanics can be accounted for without giving up the linearity of the fundamental equation.

Since our statistical model reproduces the prediction of quantum mechanics when $\sigma\rightarrow 0$, then in this limit it must necessarily violate Bell's inequality \cite{Bell beable,Kochen-Specker noncontextuality}. In other words, in the limit of $\sigma\rightarrow 0$, our statistical model is (Bell)-nonlocal. Since in this case we effectively obtain the mathematical formalism of pilot-wave theory, then the source of the non-locality can also be argued as due to the presence of a new type of potential, the last term of Eq. (\ref{HJM angular momentum}), which is called as quantum potential in pilot-wave theory. However, unlike pilot-wave theory, there is no instantaneous interaction between space-like particles for the effective velocity in our model is not the actual velocity of the particles but is an average velocity \cite{AgungDQM1}.  

An interesting question then arises for the general case when $\sigma$ is not vanishing. Since in this case our statistical model suggests a small yet finite correction to the prediction of quantum mechanics, then one can ask whether the Bell inequality is still violated or not. We expect that this issue is intimately related to the argumentation that precision test of quantum mechanics might give non-trivial limitation to the Bell non-locality test \cite{Santos review}. In this context, it is interesting to note that while there have been many experimental tests of quantum mechanics against hidden variable theories in view of Bell nonlocality and noncontextuality \cite{Genovese review}, hitherto, to our knowledge, there is no experiments which aims to directly test the precision of quantum mechanics against hidden variable theory. This might be partly  due to very few testable predictions of hidden variable theories that suggest correction to the prediction of quantum mechanics \cite{Smolin,Valentini}. 

It is then interesting to elaborate further extension of the hidden variable model discussed in the present paper. The first alternative is to give up the demand that the distribution of $\lambda$ is symmetric $P(\lambda;\hbar,\sigma)=P(-\lambda;\hbar,\sigma)$ for all values of $\sigma$ while keep assuming that the symmetry is restored in the limit $\sigma\rightarrow 0$, $\lim_{\sigma\rightarrow 0}P(\lambda;\hbar,\sigma)=P_Q(\lambda;\hbar)=\delta(\lambda-\hbar)/2+\delta(\lambda+\hbar)/2$. This will again reproduce the prediction of quantum mechanics as the first order approximation due to the smallness of $\sigma$. Since the distribution of $\lambda$ is no more symmetric, rather than using the generalized Schr\"odinger equation of (\ref{generalized Schroedinger equation angular momentum}), one should start from pair of Eqs. (\ref{FPE angular momentum}) and (\ref{HJM angular momentum}) in the case of von Neumann model. The other alternative is to drop the assumption that $\hbar$ and $\sigma$ are constant. Namely we allow one or both of them to depend on space and time: $\hbar=\hbar(q;t)$ and $\sigma=\sigma(q;t)$. Note that even if they are, assuming that the spacetime fluctuations of $\hbar$ and $\sigma$ is negligibly weak compared to the fluctuations of $\Psi(q,\lambda;t)$, then the generalized Schr\"odinger equation of (\ref{generalized Schroedinger equation angular momentum}) or (\ref{free atom Schroedinger}) are still approximately valid. 

\begin{acknowledgments} 

\end{acknowledgments}


\begin{thebibliography}{10}
\bibitem{note on constrained dynamics} Note that the classical velocity field $\underline{v}$ will depend directly on the HPF $\underline{S}$ only if the corresponding Lagrangian is not singular so that the classical velocity is expressible in term of classical momentum. Otherwise one has a constrained dynamics.
\bibitem{position measurement} We have shown in Ref. \cite{AgungDQM1} that position measurement is special in that applying the quantization scheme developed in the present paper will lead to equations exactly the same as the original classical equations. Hence, like measurement in classical mechanics, the position measurement in our hidden variable model can reveal the pre-existing value prior to the measurement.
\bibitem{high quantum number vs classical limit} Let us note that even in the standard interpretation of quantum mechanics, high quantum number does not necessarily correspond to classical limit. A famous example is given by Einstein's paradox of a single particle in a box potential when the wave function is given by the energy eigenfunction. In this case, even when the energy of the particle is very large, corresponding to high quantum number, one does not get a classical picture of particle bouncing back and forth inside the box. Also in pilot-wave theory, a wave function corresponding to a quantum state with high quantum number can still have very large quantum potential \cite{Bohm paper}. Classical limit can neither be made by going into the limit of vanishing Planck constant, $\hbar\rightarrow 0$, since the latter is fixed. In contrast to this, it is clear that in our formalism classical limit is obtained when the terms containing $\lambda$ in Eq. (\ref{fundamental equation}) is negligibly small as compared to the other terms.
\end{thebibliography}
\end{document}